\documentclass[12pt]{article}

\usepackage{authblk}
\usepackage{graphicx}
\usepackage{subcaption} 
\usepackage{epstopdf} 

\usepackage[shortlabels]{enumitem}

\usepackage[numbers,sort&compress]{natbib}

\def\eq#1{Eq.~(\ref{#1})} 
\def\eqs#1{Eqs.~(\ref{#1})} 
\def\eqn#1{(\ref{#1})} 

\graphicspath{{./figs/}}

\begin{document}
	
	\bibliographystyle{unsrt}

\title{Macroscopic Lattice Boltzmann Method for \\ Shallow Water Equations (MacLABSWE)}

\author{Jian Guo Zhou}

\affil{Department of Computing and Mathematics\\
	Manchester Metropolitan University\\
	Manchester, M1 5GD, UK\\
	J.Zhou@mmu.ac.uk}

\date{}

\maketitle

\begin{abstract}
It is well known that there are two integral steps of streaming and collision in the lattice Boltzmann method (LBM). This concept has been changed by the author's recently proposed macroscopic lattice Boltzmann method (MacLAB) to solve the Navier-Stokes equations for fluid flows.  The \mbox{MacLAB} contains streaming step only and relies on one fundamental parameter of lattice size $\delta x$, which leads to a revolutionary and precise minimal  ``Lattice'' Boltzmann method, where physical variables such as velocity and density can be retained as boundary conditions with less required storage for more accurate and efficient simulations in modelling flows using boundary condition such as Dirichlet's one.  Here, the idea for the MacLAB is further developed for solving the shallow water flow equations (MacLABSWE). This new model has all the advantages of the conventional LBM but without calculation of the particle distribution functions for determination of velocity and depth, e.g., the most efficient bounce-back scheme for no-slip boundary condition can be implemented in the similar way to the standard LBM.  The model is applied to simulate a 1D unsteady tidal flow, a 2D wind-driven flow in a dish-shaped lake and a 2D complex flow over a bump.  The results are compared with available analytical solutions and other numerical studies, demonstrating the potential and accuracy of the model. 	
\end{abstract}

\section{Introduction}
In nature, many flows have large and dominant horizontal flow characteristics compared to the vertical ones, e.g., tidal flows, waves, open channel flows, dam breaks, and atmospheric flows. Those flows are called shallow water flows and are described by the shallow water flow equations \cite{Vreugdenhil:1994}.  
As numerical solutions to the equations turn out to be a very successful tool in studying diverse flow problems encountered in engineering \cite{Vreugdenhil:1994,Alcrudo:1993,Casulli:1990,Borthwick.etc:1997,Yulistiyanto.etc:1998,Hu.etc:2000,zhou.bk.2004}, the corresponding research has received considerable attention, leading to many numerical methods ranging from finite difference method, finite element method and the Godunov type to the lattice Botzmann method. For example, Casulli \cite{Casulli:1990} proposed a semi-implicit finite difference method for the two-dimensional shallow water equations; Zhou developed a SIMPLE-like finite volume scheme to solve the shallow water equations \cite{Zhou:1995}; Alcrudo and Garcia-Navarro \cite{Alcrudo:1993} described a high resolution Godunov-type finite volume method for solution of inviscid form shallow water equations; Zhou et al. \cite{zhou.etc:2001} proposed a surface gradient method for the treatment of source terms in the shallow water equations using Godunov-type finite volume method; Zhou \cite{zhou:2002} formulated a lattice Boltzmann method for shallow water equations.

Due to the fact that the lattice Boltzmann method has been developed into a very efficient and flexible alternative numerical method in computational physics, such as nonideal fluids \cite{Swift.etc:1995}, the Brinkman equation \cite{Spid.etc:1997}, groundwater flows \cite{zhou:2007} and morphological change \cite{zhou:2014}, the study on lattice Boltzmann method for the shallow water equations has continuously been undertaken and improved: the removal of calculating the first order derivative associated with a bed slope for consistency of the lattice Boltzmann dynamics \cite{zhou:2011}, determination of theoretical relation between the coefficients in the respective local equilibrium distribution function and lattice Boltzmann equation for complex shallow water flows \cite{zhou:2013}.  This makes the development of the lattice Boltzmann method for shallow water equations (eLABSWE) to a point where it is able to produce accurate solutions to complex shallow water flow problems in an efficient way.  The method has been  
applied to several complex flow problems including large-scale practical application, demonstrating its potential, capability and accuracy in simulating shallow water flows \cite{liu.etc:2009a,liu.etc:2009b,liu.etc:2009c,liu.etc:2013}.  
  
However, the main weakness of the existing lattice Boltzmann methods for the shallow water equations is that the physical variables such as velocity and water depth cannot be applied to boundary conditions without being converted to the corresponding distribution functions.  In addition, the no-slip boundary condition cannot exactly be achieved through application of the most popular and efficient bounce-back scheme.  These drawbacks have recently been removed by Zhou \cite{Zhou.MacLAB:2018} in his proposed macroscopic lattice Boltzmann method (MacLAB) for Navier-Stokes equations to simulate fluid flows. In this paper, the MacLAB is extended to formulate the novel lattice Botlzmann method for shallow water equations (MacLABSWE).
Three numerical tests are carried out to validate the accuracy and capability of the new method. 

\section{Shallow water equations}
The 2D shallow water equations with a bed slope and a force term may be written in a tensor
notation as \cite{Zhou:1995}
\begin{equation}
\frac{\partial{h}}{\partial{t}}+
\frac{\partial{(h u_j)}}{\partial{x_j}}=0
\label{swe-con}
\end{equation}
and
\begin{eqnarray}
\frac{\partial{(hu_i)}}{\partial{t}}+
\frac{\partial{(hu_i u_j)}}{\partial{x_j}} 
=
-\frac{g}{2}\frac{\partial h^2}{\partial{x_i}}
- g h \frac{\partial z_b}{\partial x_i}
+
\nu \frac{\partial^2{(hu_i)}}{\partial{x_j^2}}
+F_i,
\label{swe-mom}
\end{eqnarray}
where $i$ and $j$ are indices and the Einstein summation convention is used, i.e. repeated indices mean a summation over the space coordinates; $x_i$ is the Cartesian coordinate; $h$ is the water depth; $t$ is the time; $u_i$ is the depth-averaged velocity component in $i^{th}$ direction; $z_b$ is the bed elevation above a datum; $g = 9.81\ m/s^2$ is the
gravitational acceleration; $\nu$ is the depth-averaged eddy viscosity;  and $F_i$ is the force term and defined as
\begin{equation}
F_i = \frac{\tau_{wi}}{\rho} - \frac{\tau_{bi}}{\rho} 
+   \Omega h u_y \delta_{ix}
-   \Omega h u_x \delta_{iy} ,
\label{SWE-force}
\end{equation}
in which $\tau_{wi}$is the wind shear stress in $i^{th}$ direction and is generally defined by
\begin{equation}
\tau_{wi}=\rho_{a} C_{w} u_{wi} \sqrt{u_{wj} u_{wj}},
\label{wind_u-stress}
\end{equation}
where $\rho_{a}=1.293$ $kg/m^{3}$ is the air density, $u_{wi}$ is the component of wind speed in $i^{th}$ direction with $C_{w}=0.0026$; and $\tau_{bi}$ is the bed shear stress in $i^{th}$ direction defined by the depth-averaged velocities as
\begin{equation}
\tau_{bi} = \rho C_b u_i \sqrt{u_ju_j}, 
\label{bed-stress}
\end{equation}
where $\rho$ is the water density and $C_b$ is the bed friction coefficient $C_z$, which is linked to Chezy coefficient as $C_b=g/C_z^2$; $\Omega$ is the Coriolis parameter for 
the effect of the earth's rotation; 
and
$\delta_{ij}$ is the Kronecker delta function,
\begin{equation}
\delta_{ij} = \left\{ \begin{array}{lr}
0,       & \hspace{9mm}  i \neq j, \\
1,       & \hspace{9mm}  i = j. 
\end{array} \right.
\label{Kronecker} \end{equation} 

\section{Macroscopic lattice Boltzmann method (MacLABSWE)}
\begin{figure}
\centering
\includegraphics[width=0.36\textwidth]{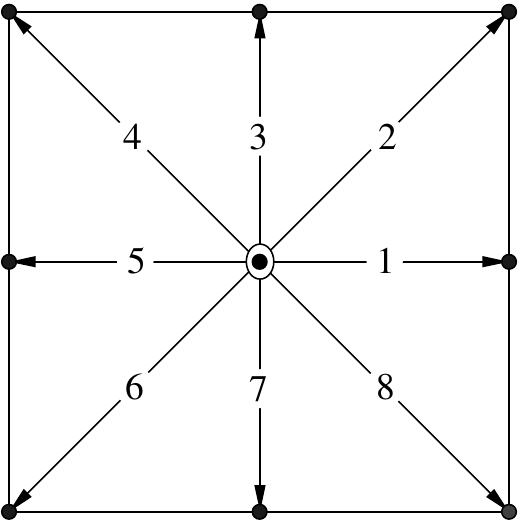}
\caption{\label{LBsquare_9.pdf} Nine-velocity square lattice (D2Q9).}
\end{figure}

The enhanced lattice Boltzmann equation for shallow water equations \eqn{swe-con} and \eqn{swe-mom}, eLABSWE, on a 2D square lattice with nine particle velocities (D2Q9) shown in Fig.~\ref{LBsquare_9.pdf} reads \cite{zhou:2011,zhou:2013} 
\begin{eqnarray}
&& \hspace{-20mm}
f_\alpha({\bf x} + {\bf e}_\alpha \delta t, t + \delta t) 
= f_
\alpha({\bf x}, t) - 
\frac{1}{\tau} [ f_\alpha ({\bf x}, t) - f_\alpha^{eq} ({\bf x}, t)  ]
\nonumber\\
&&  \hspace{15mm}
- \frac{g\overline{h}}{e^2}
C_\alpha [z_b({\bf x} + {\bf e}_\alpha \delta t)-z_b({\bf x})]
+
 \frac{\delta t}{e^2} C_\alpha e_{\alpha j}F_j,
\label{lb.1} 
\end{eqnarray}
where $f_\alpha$ is the particle distribution function; ${\bf x}$ is the space vector defined by Cartesian coordinates, i.e., ${\bf x} = (x,y)$ in 2D space; $t$ is the time;
$\delta t$ is the time step; ${\bf e}_\alpha$ is the particle velocity vector;  $e_{\alpha j}$ is the component of ${\bf e}_\alpha$ in $j^{th}$ direction;
$e=\delta x / \delta t$ is the particle speed, $\delta x$ is the lattice size;
$\tau$ is the single relaxation time \cite{Bhatnagar.etc:1954};
$C_\alpha = 1/3$ when $\alpha = 1, 3, 5, 7$ 
and $C_\alpha = 1/12$ when $\alpha = 2, 4, 6, 8$
and $f_\alpha^{eq}$ is the local equilibrium distribution function defined as
\begin{equation} 
f_\alpha^{eq} = \left\{ \begin{array}{lr}
 h \left ( 1 -\frac{5gh}{6 e^2}-\frac{2 u_i u_i }{3e^2} \right ),
&  \alpha=0, \vspace{1mm} \\
\lambda_\alpha h \left (  \frac{gh}{6 e^2}+
 \frac{e_{\alpha i}u_i }{3 e^2} +
\frac{e_{\alpha i}e_{\alpha j}u_i u_j}{2e^4} 
-\frac{u_i u_i}{6e^2}  \right ),
&  \alpha \ne 0, 
\end{array} \right.
\label{feq-full}
\end{equation}
in which $\lambda_\alpha = 1$ when $\alpha = 1, 3, 5, 7$ 
and $\lambda_\alpha = 1/4$ when $\alpha = 2, 4, 6, 8$; and $\overline{h} = 0.5 [h({\bf x} + {\bf e}_\alpha \delta t, t+\delta t) + h({\bf x}, t)]$.
The physical variables of water depth and velocity can be calculated as
\begin{equation}
h = \sum_\alpha f_\alpha  ,
\label{fea-0-h}
\end{equation}
and
\begin{equation}
u_i = \frac{1}{h} 
\sum_\alpha e_{\alpha i} f_\alpha  .
\label{fea-0-v}
\end{equation}

To formulate a new macroscopic lattice Boltzmann method for the shallow water equations through the macroscopic physical variables of velocity and water depth without calculating  distribution functions, \eq{lb.1} is rewritten as
\begin{eqnarray}
f_\alpha({\bf x}, t) 
&=&
f_\alpha({\bf x} - {\bf e}_\alpha \delta t, t - \delta t) 
- \frac{1}{\tau} [ f_\alpha({\bf x} - {\bf e}_\alpha \delta t, t - \delta t) 
\nonumber\\
&-& f_\alpha^{eq}({\bf x} - {\bf e}_\alpha \delta t, t - \delta t) ]
\nonumber\\  \vspace{1cm}
&-&
\frac{g \overline{h}}{e^2}
C_\alpha [z_b({\bf x}) -  z_b({\bf x} - {\bf e}_\alpha \delta t)]
+ \frac{\delta t}{e^2} C_\alpha e_{\alpha j} F_j .
\label{lb.Mac.1} 
\end{eqnarray}
Following Zhou's idea in MacLAB \cite{Zhou.MacLAB:2018}, setting $\tau = 1$ in the above equation leads to
\begin{eqnarray}
f_\alpha ({\bf x}, t)
& = &
f_\alpha^{eq} ({\bf x} - {\bf e}_\alpha \delta t, t - \delta t)
\nonumber\\
&-&
\frac{g \overline{h}}{e^2}
C_\alpha [z_b({\bf x}) -  z_b({\bf x} - {\bf e}_\alpha \delta t)]
+ \frac{\delta t}{e^2} C_\alpha e_{\alpha j} F_j.
\label{Mac.lb.3} 
\end{eqnarray}
Taking $\sum$ \eq{Mac.lb.3} and $\sum e_{\alpha i}$\eq{Mac.lb.3} yields
\begin{eqnarray}
\hspace{-9mm}
\sum f_\alpha ({\bf x}, t)
& = &
\sum f_\alpha^{eq} ({\bf x} - {\bf e}_\alpha \delta t, t - \delta t)
\nonumber\\
&-& 
\frac{g}{e^2}
\sum C_\alpha \overline{h} [z_b({\bf x}) -  z_b({\bf x} - {\bf e}_\alpha \delta t)]
+  \frac{\delta t}{e^2} \sum  C_\alpha e_{\alpha j}F_j,
\label{Mac.lb.4} 
\end{eqnarray}
and
\begin{eqnarray}
\sum e_{\alpha i} f_\alpha ({\bf x}, t)
& = &
\sum e_{\alpha i} f_\alpha^{eq} ({\bf x} - {\bf e}_\alpha \delta t, t - \delta t)
+ \frac{\delta t}{e^2} \sum  C_\alpha e_{\alpha i} e_{\alpha j}F_j
\nonumber\\
&-& 
\frac{g}{e^2}
\sum C_\alpha e_{\alpha i} \overline{h} [z_b({\bf x}) -  z_b({\bf x} - {\bf e}_\alpha \delta t)] .
\label{Mac.lb.5} 
\end{eqnarray}
As $\sum f_\alpha ({\bf x}, t) = h ({\bf x}, t) $  and $\sum e_{\alpha i} f_\alpha ({\bf x}, t) = h ({\bf x}, t) u_i({\bf x}, t) $ due to the requirement for the conservation of mass and momentum in the lattice Botlzmann dynamics, the above two equations become
\begin{eqnarray}
h ({\bf x}, t)
& = &
\sum f_\alpha^{eq} ({\bf x} - {\bf e}_\alpha \delta t, t - \delta t)
\nonumber\\
&-& 
\frac{g}{e^2}
\sum C_\alpha \overline{h} [z_b({\bf x}) -  z_b({\bf x} - {\bf e}_\alpha \delta t)]
+  \frac{\delta t}{e^2} \sum  C_\alpha e_{\alpha j}F_j
\label{Mac.lb.6} 
\end{eqnarray}
and
\begin{eqnarray}
h ({\bf x}, t) u_i ({\bf x}, t)
& = &
\sum e_{\alpha i} f_\alpha^{eq} ({\bf x} - {\bf e}_\alpha \delta t, t - \delta t)
+ \frac{\delta t}{e^2} \sum  C_\alpha e_{\alpha i} e_{\alpha j}F_j
\nonumber\\
&-& 
\frac{g}{e^2}
\sum C_\alpha e_{\alpha i} \overline{h} [z_b({\bf x}) -  z_b({\bf x} - {\bf e}_\alpha \delta t)] .
\label{Mac.lb.7} 
\end{eqnarray}
According to the centred scheme \cite{zhou.bk.2004,Zhou.AxLBM:2011} 
the force term $F_j$ can be evaluated at the midpoint between $({\bf x} - {\bf e}_\alpha \delta t, t - \delta t)$ and $({\bf x}, t)$ as 
\begin{equation}
F_j = F_j \left ( {\bf x} - \frac{1}{2}{\bf e}_\alpha \delta t, t - \frac{1}{2}\delta t \right ).
\label{force-CS}
\end{equation} 
It can be seen from \eqs{Mac.lb.6} and \eqn{Mac.lb.7} that the water depth and velocity can be determined using the macroscopic physical variables through the local equilibrium distribution function without calculating the distribution function from \eq{lb.1} that is required in \eqs{fea-0-h} and \eqn{fea-0-v} for determination of the depth and velocity.  These equations form the macroscopic lattice Boltzmann method for shallow water equations (MacLABSWE).  It shows through the recovery procedure in Appendix that the eddy viscosity $\nu$ in the absence of collision step can be naturally taken into account using the particle speed $e$ from
\begin{equation}
e = 6 {\nu}/{ \delta x },
\label{mlb-viscosity}
\end{equation}
instead of $e=\delta x/\delta t$ to calculate the local equilibrium distribution function $f_\alpha^{eq}$ from \eq{feq-full}.  Apparently, after a lattice size $\delta x$ is chosen, the model is ready to simulate a flow with an eddy viscosity $\nu$ because $(x_j - e_{\alpha j} \delta t)$ stands for a neighbouring lattice point; $f_\alpha^{eq}$ at time of $(t-\delta t)$ represents its known quantity at the current time; and the particle speed $e$ is determined from \eq{mlb-viscosity}  for use in computation of $f_\alpha^{eq}$. In addition, the time step $\delta t$ is no longer an independent parameter but is calculated as $\delta t = \delta x/e$, which is used in simulations of unsteady flows.  Consequently, only the lattice  size $\delta x$ is required in the MacLABSWE for simulation of shallow water flows, bringing the eLABSWE into a precise ``Lattice'' Boltzmann method for shallow water flows. This enables the model to become an automatic simulator without tuning other simulation parameters, making it possible and easy to model a large flow system when a super-fast computer such as a quantum computer becomes available in the future. 

The method is unconditionally stable as it shares the same valid condition as that for $f_\alpha^{eq}$, or the Mack number $M=U_c/e$ is much smaller than 1, in which $U_c$ is a characteristic flow speed. The Mack number can also be expressed as a lattice Reynolds number of $R_{le}=U_c \delta x/ \nu$ via \eq{mlb-viscosity}. In practical simulations, it is found that the model is stable if $R_{le}=U_m \delta x/ \nu < 1$ where $U_m$ is the maximum flow speed and is used as the characteristic flow speed. The main features of the MacLABSWE are that there is no collision operator and only macroscopic physical variables such as depth and velocity are required, which are directly retained as boundary conditions with a minimum memory requirement. At the same time, the most efficeint bounce-back scheme can be implemented as that in the standard lattice Botlzmann method if it is required, e.g., if the water depth is unknown and no-slip boundary condition is applied at south boundary for a straight channel, $f_2^{eq},\  f_3^{eq}, \ f_4^{eq}$ in \eq{Mac.lb.6} are unknown and they can be determined as $f_2^{eq}=f_6^{eq},\ f_3^{eq}=f_7^{eq},\ f_4^{eq}=f_8^{eq}$ using the bounce-back scheme, after which the water depth can be determined from \eq{Mac.lb.6} and in this case \eq{Mac.lb.7} is no longer required for calculation of velocity as the initial zero velocity will retain as no-slip boundary condition there.
The simulation procedure for MacLABSWE is
\begin{enumerate}[(a)]
	\item Initialise water depth and velocity,
	\item \label{step2} Choose the lattice size $\delta x$ and determine the particle speed $e$ from \mbox{\eq{mlb-viscosity}}, 
	\item Calculate $f_\alpha^{eq}$ from \eq{feq-full} using depth and velocity, 
	\item Update the depth and velocity using \eqs{Mac.lb.6} and \eqn{Mac.lb.7}, 
	\item Apply the boundary conditions if necessary, and repeat Step \ref{step2} until a solution is reached.
\end{enumerate}
The only limitation of the described model is that, for small eddy viscosity or high speed flow, the chosen lattice size after satisfying $R_{le} < 1$ may turn out to generate very large lattice points (Lattice points, e.g., for one dimension with length of $L$ is calculated as $N_L=L/\delta x$ and $N_L$ is the lattice points); if the total lattice points is too big such that the demanding computations is beyond the current power of a computer, the simulation cannot be carried out.  Such difficulties may be solved or relaxed through parallel computing using computer techniques such as GPU processors and multiple servers, and will largely or completely removed using quantum computing when a quantum computer becomes available.

\section{Validation}

In order to verify the described model, 
three numerical tests are presented.  
The SI Units are used for the physical variables in the following numerical simulations. 

\subsection{1D tidal flow}
\begin{figure}
\centering
\includegraphics[width=0.82\textwidth]{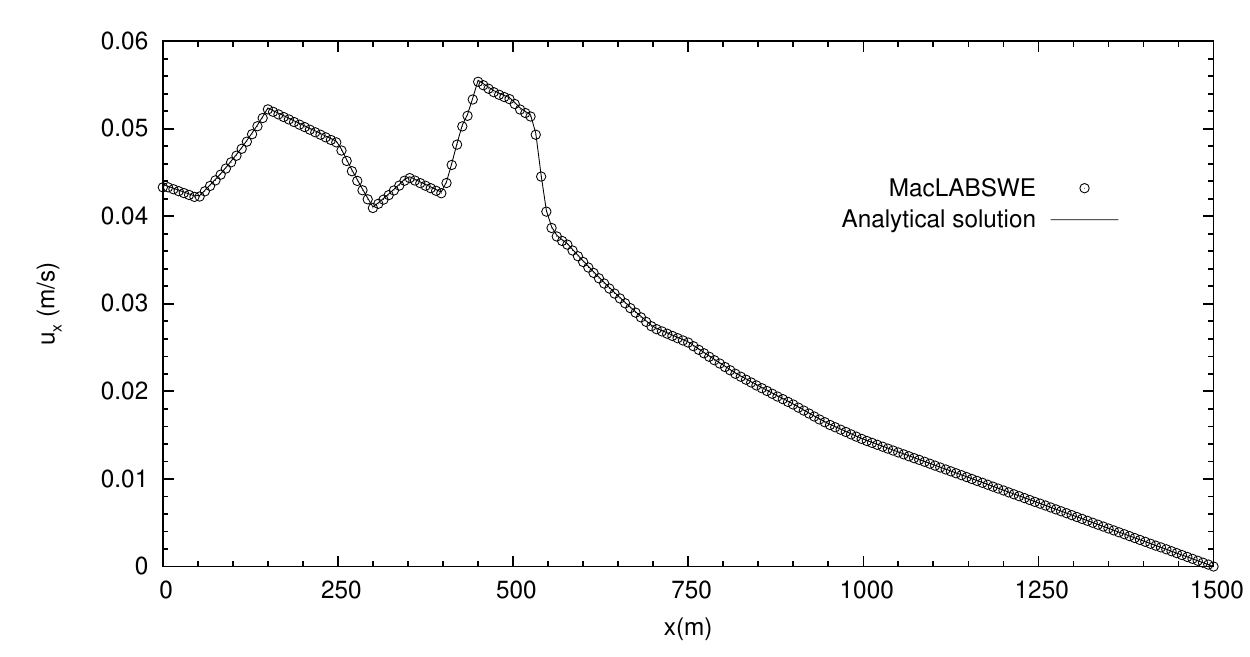}
\caption{Comparison of velocity at $t=10,800\ s$ when flow is in the half-risen tide with maximum positive velocities for 1D tidal flow.}
\label{1D-tidal-com+u.pdf}
\end{figure}

\begin{figure}
\centering
\includegraphics[width=0.82\textwidth]{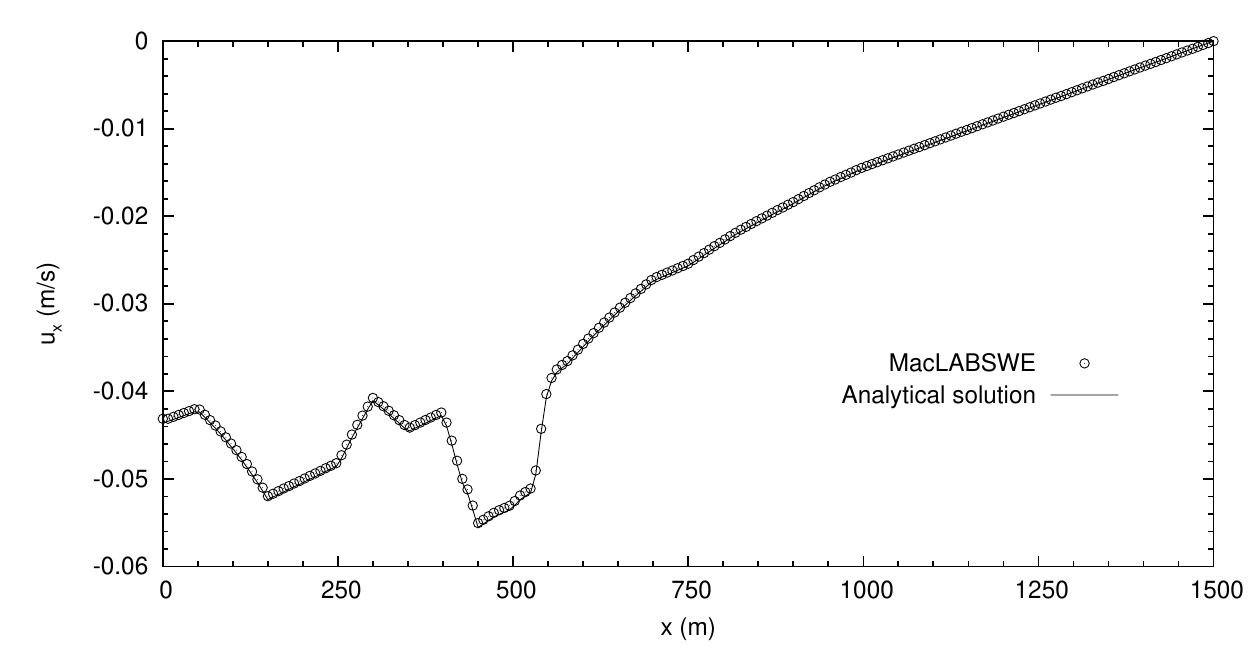}
\caption{Comparison of velocity at $t=32,400\ s$ when the flow is in the half-ebb tide with maximum negative velocities for 1D tidal flow.}
\label{1D-tidal-com-u.pdf}
\end{figure}

First of all, a tidal flow over an irregular bed is predicted, 
which is a common flow problem in coastal engineering.
The bed is defined with data listed in Table~\ref{tidal-irre.zb}. 
\begin{table}[h]
\caption{Bed elevation $z_b$ for irregular bed.}
\centering
\begin{tabular}{c c c c c c c c c c c}
\hline
$x (m)$   & 0 & 50 & 100 & 150 & 250 & 300 & 350 & 400 & 425 & 435 \\
$z_b (m)$ & 0 & 0  & 2.5 & 5   & 5   & 3   & 5   & 5   & 7.5 & 8    \\
\hline
$x (m)$   & 450 & 475 & 500 & 505 & 530 & 550 & 565 & 575 & 600 & 650 \\
$z_b (m)$ & 9   & 9  & 9.1 & 9 & 9   & 6   & 5.5 & 5.5 & 5   & 4  \\
\hline
$x (m)$   & 700 & 750 & 800 & 820 & 900 & 950 & 1000 & 1500 &  &  \\
$z_b (m)$ & 3   & 3   & 2.3 & 2   & 1.2 & 0.4 & 0    & 0    &  &  \\
\hline
\end{tabular}
\label{tidal-irre.zb}
\end{table}
Here we consider a 1D problem with the initial and boundary conditions of
\begin{equation}
h(x,0)=16-z_b(x),
\label{veri-Ini-Bc.1r}
\end{equation}
\begin{equation}
u_x(x,0)=0
\label{veri-Ini-Bc.2r}
\end{equation}
and
\begin{equation}
h(0,t)=20-4 \sin \left [\pi \left ( \frac{4t}{86,400}+\frac{1}{2}
 \right ) \right ],
\label{veri-Ini-Bc.3r}
\end{equation}
\begin{equation}
u_x(1500,t)=0.
\label{veri-Ini-Bc.4r}
\end{equation}
In the simulation, $\delta x=7.5 \ m$ or $200$ lattices are used with 
eddy viscosity of $\nu = 31.25\ m^2/s$ for same computational parameters used in \cite{zhou:2011}.  This is an unsteady flow.
Two numerical results at $t=10,800 \ s$ and $t=32,400 \ s$
corresponding to the half-risen tidal flow with maximum
positive velocities and to the half-ebb tidal
flow with maximum negative velocities
are compared with the  
analytical solutions \cite{Bermudez.etc:1994} and depicted 
in Figs.~\ref{1D-tidal-com+u.pdf} and \ref{1D-tidal-com-u.pdf}, respectively.
The maximum relative errors are less than 0.005\% for the water level, less than 0.05\% for velocity larger than 0.002 $m/s$, and less than 0.3\% for smaller velocity, revealing excellent agreements.  

\subsection{2D wind-driven circulation}
\begin{figure}
\centering
\includegraphics[width=0.75\textwidth]{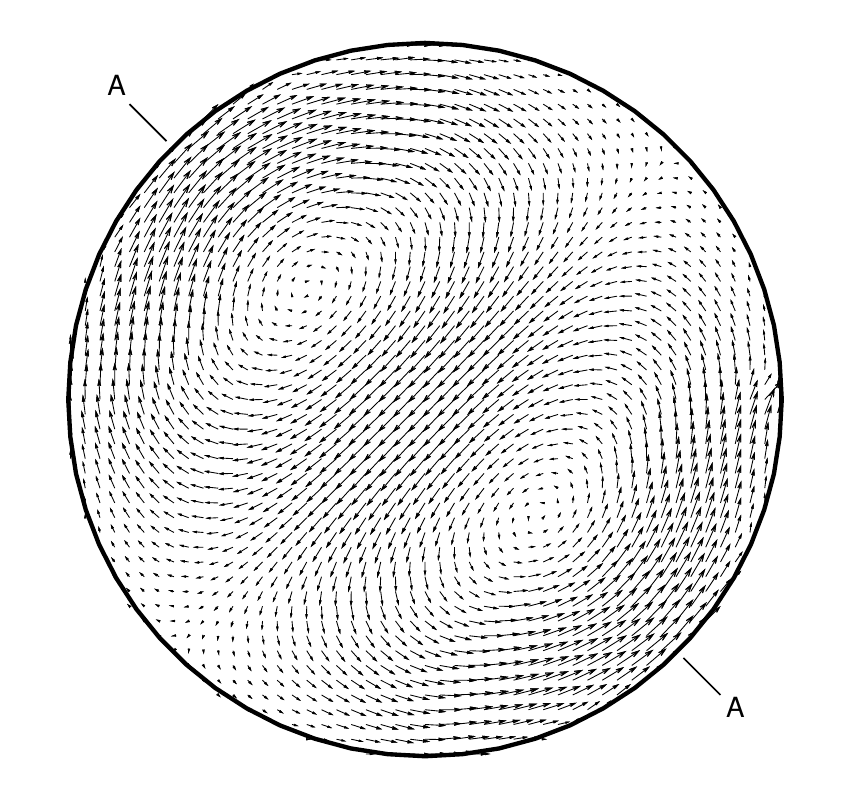}
\caption{Flow field for wind-driven flow, showing well-developed
counter-rotating gyres with flow in the deeper water against the 
direction of the wind.}
\label{2D-wind-vec.pdf}
\end{figure}

\begin{figure}
\centering
\includegraphics[width=0.8\textwidth]{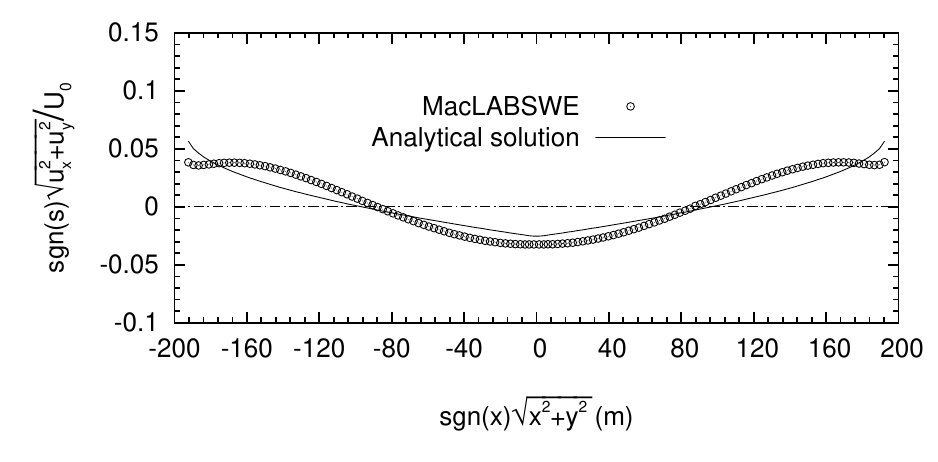}
\caption{Comparison of the resultant velocities along Cross-section A-A (see Fig.~\ref{2D-wind-vec.pdf}) with the analytical solution \cite{Kranenburg:1992}, where
$U_0 = 0.129$ and $\ s=u_x+u_y$.}
\label{2D-wind-Com.pdf}
\end{figure}

Secondly, we consider a wind-driven circulation in a lake,
which may generate a complex flow phenomenon depending on the bed topography
of a lake.  In this test, a uniform wind shear stress is applied
to the shallow water in a circular basin with the bed topography
 defined by the still water depth $H$,
\begin{equation}
H(x,y)=\frac{1}{1.3}\left ( \frac{1}{2}+\sqrt{\frac{1}{2}- \frac{\sqrt{x^2+y^2}}{386.4}} \right ),
\label{wind_bed}
\end{equation}
from which, the bed level can be determined as
$ z_b(x,y)=H(0,0)-H(x,y) $.
The same dish-shaped basin is also used by Rogers et al. \cite{Rogers.etc:2001} to test
a Godunov-type method. Initially, the water in the basin is still and then a uniform wind speed of $u_{w}=5\ m/s$ blows from southwest to northeast, at which wind shear stress is calculated from \eq{wind_u-stress}.
Its steady flow consists of two relatively strong
counter-rotating gyres with flow in the deeper water against the 
direction of the wind, exhibiting complex flow phenomenon.    
In the numerical computation, $\delta x = 2$ or $200 \times 200$ lattices are used with eddy viscosity of $\nu = 5.33\ m^2/s$.
After the steady solution is obtained, the flow field is shown in Fig.~\ref{2D-wind-vec.pdf} and
the normalised resultant velocities at cross section $A-A$ 
are compared with the analytical solution \cite{Kranenburg:1992} in Fig.~\ref{2D-wind-Com.pdf}, exhibiting similar agreement 
to that by Zhou \cite{zhou:2013} for the same test.
Although there is discrepancy between the numerical prediction and the analytical solution, 
such agreement is reasonable due to the fact that the assumptions of both the rigid-lid approximation for 
the water surface and a parabolic distribution for the eddy viscosity 
were used in deriving the analytical solution. 

\subsection{Flow over a 2D hump}
\begin{figure}
	\centering
	\includegraphics[width=\columnwidth]{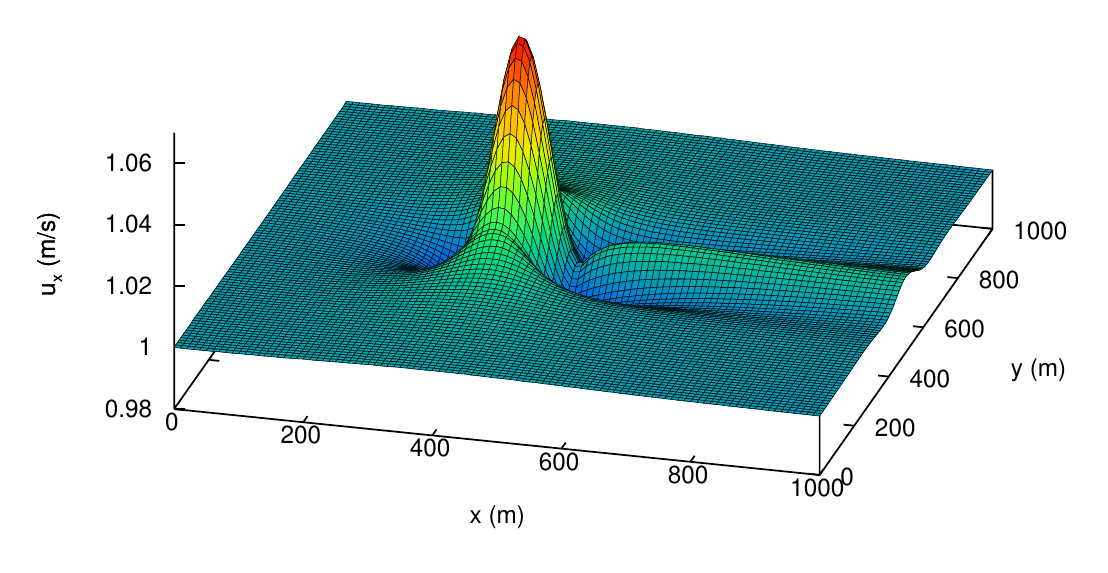}
	\caption{Velocity $u_x$ distribution for a steady flow over a 2D bump.}
	\label{2D-hump-U-A6-B24.ps}
\end{figure}

\begin{figure}
	\centering
	\includegraphics[width=\columnwidth]{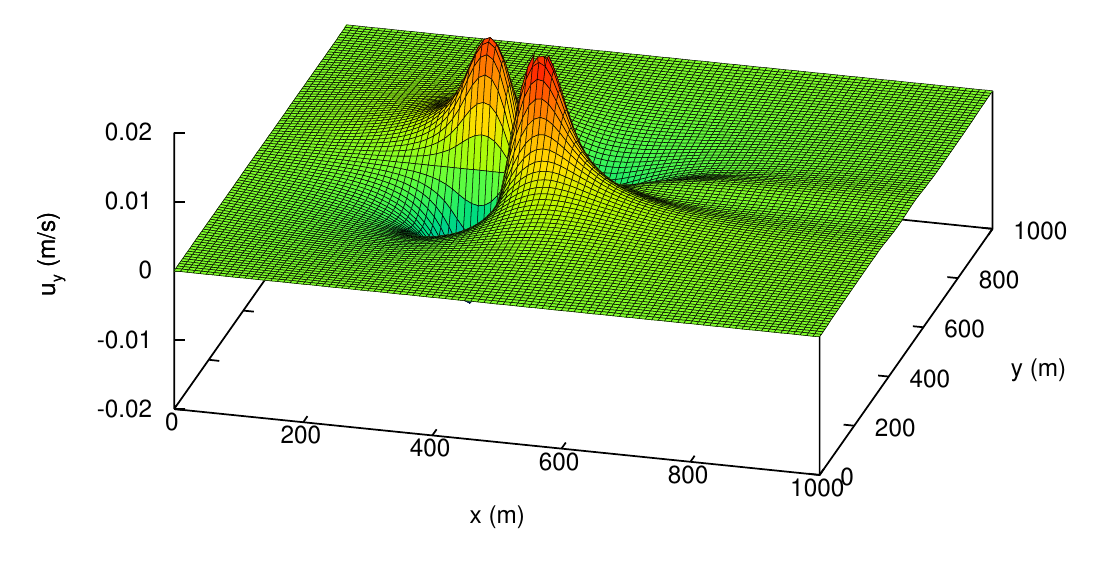}
	\caption{Velocity $u_y$ distribution for a steady flow over a 2D bump.}
	\label{2D-hump-V-A6-B24.ps}
\end{figure}

Finally, a steady shallow water flow over a 2D hump is investigated.  The 2D hump is defined as
\begin{equation}
z_b(x,y)=\left\{ \begin{array}{lr}
\psi(x,y) , & \mbox{if } (x,y)\in \Omega, \\
0, & \mbox{otherwise},
\end{array} \right.
\label{2D sand bed}
\end{equation}
where $\Omega = [300,500] \times [400,600]$ and
\begin{equation}
\psi(x,y)=\sin^2 \left ( \frac{\pi(x-300)}{200} \right ) \sin^2 \left ( \frac{\pi(y-400)}{200} \right ).
\label{2D sand bedAdd}
\end{equation}
The flow conditions are: discharge per unit width is $q=10\ m^2/s$; water depth is $h=10\ m$ at the outflow boundary and the channel is $1000\ m$ long and $1000\ m$ wide. This is the same test as that used by researchers  in validation of numerical methods \cite{Hudson.etc:2005,Huang.etc:2010,Benkhaldoun.etc:2010} for sediment transport under shallow water flows.  Here only steady flow over the fixed bed without sediment transport is simulated as prediction of correct flow plays an essential role in determination of bed evolution, and hence it is a suitable test for the proposed scheme.  We use $\delta x = 5$ or $200 \times 200$ lattices in the simulation.  After the steady solution is obtained, the velocities $u_x$ and $u_y$ are shown in Figs.~\ref{2D-hump-U-A6-B24.ps} and \ref{2D-hump-V-A6-B24.ps}, respectively, demonstrating good agreements with those obtained using high-resolution Godunov-type numerical methods \cite{Hudson.etc:2005,Huang.etc:2010,Benkhaldoun.etc:2010}. 

\section{Conclusions}
The paper presents a novel macroscopic lattice Boltzmann method for shallow water equations (MacLABSWE).  Only streaming step is required in the model.  This changes the stardard view of two integral steps of streaming and collision in the lattice Boltzmann method.  The method is unconditionally stable.  The physical variables can be directly applied as boundary condition without coverting them to their corresponding distribution functions.  This greatly simplifies the procedure and needs less storage of computer. The MacLABSWE preserves the simple arithmetic calculations of the lattice Boltzmann method at the full advantages of the lattice Boltzann method. The most efficient bounce-back scheme can be applied straightforward if it is required.  Steady and unsteady numerical tests have shown that the
method can provide accurate solutions, making the MacLABSWE an ideal model for simulating shallow water flows.

\section*{Appendix: Recovery of shallow water equations}
\label{recovery}
As the MacLABSWE is the special case where $\tau = 1$ in \eq{lb.Mac.1}, without loss of generality, we can show how to recover the shallow water eqautions \eqn{swe-con} and \eqn{swe-mom} from it. For this 
we take a Taylor expansion to the terms on the right-hand side of \eq{lb.Mac.1}, $f_\alpha({\bf x} - {\bf e}_\alpha \delta t, t - \delta t) $ and $ f_\alpha^{eq}({\bf x} - {\bf e}_\alpha \delta t, t - \delta t)$,  in time and space
at point $({\bf x}, t)$, and have
\begin{eqnarray}
f_\alpha({\bf x} - {\bf e}_\alpha \delta t, t - \delta t) 
& = & f_\alpha -
\delta t \left ( \frac{\partial}{\partial t}  + e_{\alpha j}
\frac{\partial}{\partial x_j} \right ) f_\alpha
\nonumber \\
& + &
\frac{1}{2} \delta t^2 \left ( \frac{\partial}{\partial t}+e_{\alpha j}
\frac{\partial}{\partial x_j} \right )^2 f_\alpha 
+ {\cal O} (\delta t^3)
\label{ex.lb.3} 
\end{eqnarray}
and
\begin{eqnarray}
f_\alpha^{eq}({\bf x} - {\bf e}_\alpha \delta t, t - \delta t) 
& = & f_\alpha^{eq} -
\delta t \left ( \frac{\partial}{\partial t}  + e_{\alpha j}
\frac{\partial}{\partial x_j} \right ) f_\alpha^{eq}
\nonumber \\
& + &
\frac{1}{2} \delta t^2 \left ( \frac{\partial}{\partial t}+e_{\alpha j}
\frac{\partial}{\partial x_j} \right )^2 f_\alpha^{eq} 
+ {\cal O} (\delta t^3) .
\label{ex.lb.4} 
\end{eqnarray}
According to the Chapman-Enskog analysis,  
$f_\alpha$ can be expanded in a series of $\delta t$,
\begin{equation}
f_\alpha = f_\alpha^{(0)} + \delta t f_\alpha^{(1)} +
\delta t^2 f_\alpha^{(2)} + {\cal O} (\delta t^3).
\label{fa-ex.1} \end{equation}
\eq{force-CS} can be written, via a Taylor expansion, as
\begin{eqnarray}
F_j \left ( {\bf x} - \frac{1}{2}{\bf e}_\alpha \delta t, t - \frac{1}{2}\delta t \right ) = F_j -
\frac{\delta t}{2} \left ( \frac{\partial}{\partial t}  + e_{\alpha j}
\frac{\partial}{\partial x_j} \right ) F_j  + {\cal O} (\delta t^2) .
\label{ex.lb.6} 
\end{eqnarray}
The forth term on the right hand side of \eq{lb.Mac.1} can also be expressed via the Taylor expansion,
\begin{eqnarray}
\hspace{-3mm}
\frac{g C_\alpha}{e^2} \left [h - \frac{\delta t}{2} \left ( \frac{\partial h}{\partial t}+e_{\alpha j}
\frac{\partial h}{\partial x_j} \right ) \right ] 
\hspace{-1mm}
\left ( \delta t  e_{\alpha j} \frac{\partial z_b}{\partial x_j}  
- \frac{\delta t^2 }{2} e_{\alpha i} e_{\alpha j} \frac{\partial^2 z_b}{\partial x_i \partial x_j}
\right ) 
\hspace{-1mm}
+ {\cal O} (\delta t^3).
\label{lb.zb} \end{eqnarray}
After substitution of \eqs{ex.lb.3} - \eqn{lb.zb} into \eq{lb.Mac.1}, we have the expressions to order $\delta t^0$
\begin{equation}
f_\alpha^{(0)} = f_\alpha^{eq},
\label{order:e-0} \end{equation}
to order $\delta t$
\begin{equation}
\left (\frac{\partial}{\partial t}+e_{\alpha j}
\frac{\partial}{\partial x_j} \right )  f_\alpha^{(0)}
= 
-\frac{f_\alpha^{(1)}}{\tau}
- \frac{gh C_\alpha e_{\alpha j}}{e^2}  \frac{\partial z_b}{\partial x_j}
+  \frac{C_\alpha e_{\alpha j} F_j}{e^2} ,
\label{Chpman-Enskog.1} \end{equation}
and to order $\delta t^2$ as
\begin{eqnarray}
&& \hspace{-13mm}
\left ( 1-\frac{1}{\tau}  \right )
\left (\frac{\partial}{\partial t} \right .
+
\left .
e_{\alpha j}
\frac{\partial}{\partial x_j} \right ) f_\alpha^{(1)}
- \frac{1}{2} \left( \frac{\partial}{\partial t}+e_{\alpha j}
\frac{\partial}{\partial x_j} \right )^2 f_\alpha^{(0)}
= 
\nonumber\\ 
&&  \hspace{13mm}
-
\frac{1}{\tau} f_\alpha^{(2)}
+ 
\frac{C_\alpha g e_{\alpha j}}{2e^2} \left ( \frac{\partial h}{\partial t}+e_{\alpha i}
\frac{\partial h}{\partial x_i} \right ) \frac{\partial z_b}{\partial x_j}
\nonumber\\
&& \hspace{13mm}
+
\frac{gh C_\alpha e_{\alpha i} e_{\alpha j}}{2e^2}  \frac{\partial^2 z_b}{\partial x_i \partial x_j}
-
\frac{C_\alpha e_{\alpha j}}{2e^2} \left ( \frac{\partial F_j}{\partial t}+e_{\alpha i}
\frac{\partial F_j}{\partial x_i} \right ) .
\label{Chpman-Enskog.2} \end{eqnarray}
Substitution of \eq{Chpman-Enskog.1} into
\eq{Chpman-Enskog.2} gives
\begin{equation}
\left (1- \frac{1}{2\tau} \right ) \left ( \frac{\partial}{\partial t}+e_{\alpha j}
\frac{\partial}{\partial x_j} \right )
f_\alpha^{(1)}
=-\frac{1}{\tau} f_\alpha^{(2)}.
\label{Chpman-Enskog.4} \end{equation}
Taking $\sum$ [\eqn{Chpman-Enskog.1} +
$\delta t  \times$ \eqn{Chpman-Enskog.4}] about $\alpha$ provides
\begin{equation}
\frac{\partial}{\partial t} \sum_\alpha f_\alpha^{(0)}
+ \frac{\partial}{\partial x_j} \sum_\alpha e_{\alpha j} f_\alpha^{(0)}
=0.
\label{conuity-eq.1}
\end{equation}
Evaluation of the terms in the above equation using 
\eq{feq-full}
results in the second-order accurate continuity equation \eqn{swe-con}.

Taking $\sum$ $e_{\alpha i}$ [\eqn{Chpman-Enskog.1} +
$\delta t  \times$ \eqn{Chpman-Enskog.4}] about $\alpha$ yields
\begin{eqnarray}
&& \hspace{-9mm}
\frac{\partial}{\partial t} \sum_\alpha e_{\alpha i} f_\alpha^{(0)}
+  \frac{\partial}{\partial x_j}
\sum_\alpha e_{\alpha i} e_{\alpha j} f_\alpha^{(0)} +
\nonumber \\
&& \hspace{6mm}
\delta t (1- \frac{1}{2\tau}) \frac{\partial}{\partial x_j}
\sum_\alpha e_{\alpha i} e_{\alpha j} f_\alpha^{(1)}
= - gh  \frac{\partial z_b}{\partial x_i} + F_i.
\label{Momum-eq.2} \end{eqnarray}
After the terms are simplified with \eq{feq-full} and some algebra,
the above equation becomes the momentum equation \eqn{swe-mom}, which is second-order accurate, where the eddy viscosity $\nu$ is defined by
\begin{equation}
\nu = \frac{e^2 \delta t}{6} (2\tau -1) .
\label{viscosity}
\end{equation}
As the above general derivation is carried out for a constant of $\tau$, setting $\tau=1$ also recovers the shallow water equations. In this case, \eq{viscosity} becomes \eq{mlb-viscosity}.

It must be pointed out that (a) the implicitness related to $\overline h$ can be eliminated by using the method
by He et al. \cite{he.etc:1998}; (b) alternatively, the following semi-implicit form,
\begin{equation}
\overline{h} = 0.5 [h({\bf x}, t) + h({\bf x} - {\bf e}_\alpha \delta t, t) ],
\label{h-bar}
\end{equation}
can be used, which is simple and demonstrated to produce accurate solutions, and
hence it is preferred in practice.


\end{document}